\documentclass[aps,prl,twocolumn,superscriptaddress,showkeys]{revtex4-1}

\usepackage{amsmath}
\usepackage{amssymb}
\usepackage{graphicx}
\usepackage{color}
\usepackage{soul}

\begin{document}

\title{Blueshift of the surface plasmon resonance in silver nanoparticles studied with EELS}
\date{\today}

\author{S\o ren Raza,$^{1,2,*}$ Nicolas Stenger,$^{1,3,*}$ Shima Kadkhodazadeh,$^2$ S\o ren V. Fischer,$^4$ Natalie~Kostesha,$^4$ Antti-Pekka Jauho,$^{4,3}$ Andrew Burrows,$^2$ Martijn Wubs,$^{1}$ and N. Asger Mortensen$^{1,3,\dag}$\\{\scriptsize ~\\
 $^1$Department of Photonics Engineering, Technical University of Denmark, DK-2800 Kgs. Lyngby, Denmark\\$^2$Center for Electron Nanoscopy, Technical University of Denmark, DK-2800 Kgs. Lyngby, Denmark\\$^3$Center for Nanostructured Graphene (CNG), Technical University of Denmark, DK-2800 Kgs. Lyngby, Denmark\\$^4$Department of Micro and Nanotechnology, Technical University of Denmark, DK-2800 Kgs. Lyngby, Denmark\\~\\$^*$Both authors contributed equally\\$^\dag$Corresponding author email: asger@mailaps.org}}

\begin{abstract}
We study the surface plasmon (SP) resonance energy of isolated spherical Ag nanoparticles dispersed on a silicon nitride substrate in the diameter range 3.5-26~nm with monochromated electron energy-loss spectroscopy. A significant blueshift of the SP resonance energy of 0.5~eV is measured when the particle size decreases from 26 down to 3.5~nm. We interpret the observed blueshift using three models for a metallic sphere embedded in homogeneous background material: a classical Drude model with a homogeneous electron density profile in the metal, a semiclassical model corrected for an inhomogeneous electron density associated with quantum confinement, and a semiclassical nonlocal hydrodynamic description of the electron density. We find that the latter two models provide a qualitative explanation for the observed blueshift, but the theoretical predictions show smaller blueshifts than observed experimentally.
\end{abstract}

\maketitle

\section{Introduction}

Surface plasmons are collective excitations of the electron gas in metallic structures at the metal/dielectric interface~\cite{Maier:2007}. The ability to concentrate light with SPs~\cite{Schuller:2010} and to enhance light-matter interaction on a subwavelength scale enables few- and even single-molecule spectroscopy when the size of the metallic structures is decreased to a few nanometer~\cite{Kneipp:2007}.
These collective excitations are usually well-described by the classical Drude model for nanoparticles with dimensions of tens of nanometer and larger~\cite{Maier:2007}. In the quasistatic limit, \textit{i.e.} when the wavelength of the exciting electromagnetic wave considerably exceeds the dimensions of the structure, the local-response Drude theory predicts that the resonance energy of localized SPs is independent of the size of the nanostructure~\cite{Wang:2006}, and that the field enhancement created in the gap between two metallic nanostructures diverges for vanishing gap size~\cite{Romero:2006}. These predictions are however in conflict both with earlier~\cite{Kreibig:1985,Charle:1989,Ouyang:1992,Berciaud:2005} and with more recent experimental results, which have shown a size dependency of the localized SP resonance in noble metal nanoparticles in the size range of 1-10 nm~\cite{Scholl:2012} and pronounced deviations for dimer geometries~\cite{Ciraci:2012, Kern:2012}.

This dependence of the SP resonance on the size of noble metal nanostructures is believed to be a signature of quantum properties of the free-electron gas. With decreasing sizes of the nanoparticles, the quantum wave nature of the electrons is theoretically expected to manifest itself in the optical response due to the effects of quantum confinement~\cite{Genzel:1975,Kraus:1983,Halperin:1986,Keller:1993,Ozturk:2011a}, quantum tunneling~\cite{Zuloaga:2009,Mao:2009a,Ozturk:2011a,Esteban:2012}, as well as nonlocal response~\cite{Ljungbert:1985,Garcia-de-Abajo:2008,David:2012,Aizpurua:2008,Raza:2011,Toscano:2012,Fernandez-Dominguez:2012}. Nonlocal effects are a direct consequence of the inhomogeneity of the electron gas, which arises due to the quantum wave nature and the many-body properties of the electron gas.

The recent developments in analytical scanning transmission electron microscopes (STEM) equipped with a monochromator and electron energy-loss spectroscopy (EELS)~\cite{GarciadeAbajo:2010} give the possibility of accessing the near-field energy distribution of the plasmon resonance of individual nanoparticles on a subnanometer scale with an energy resolution better than 0.2~eV. This method has been used for the imaging of surface plasmons in many different metallic nanostructures~\cite{Nelayah:2007,Koh:2009,Nicoletti:2011,Koh:2011,Scholl:2012}.
With STEM EELS it is possible to correlate the structural and chemical information on the nanometer scale, such as the shape and the presence of organic ligands, with the spectral information of the SP resonance of single isolated nanoparticles. STEM EELS is thus perfectly suited to probe and access plasmonic nanostructures and SP resonances at length scales where quantum mechanics is anticipated to become important.

In this paper we report the experimental study of the SP resonance of chemically grown single Ag nanoparticles dispersed on 10~nm thick $\text{Si}_3\text{N}_4$ membranes with STEM EELS. Our measurements present a significant blueshift of the SP resonance energy from 3.2 to 3.7~eV for particle diameters ranging from 26 down to 3.5~nm. Our results also confirm very recent experiments made with Ag nanoparticles on different substrates using different STEM operating conditions~\cite{Scholl:2012}, thereby strengthening the interpretation that the blueshift is predominantly associated with the tight confinement of the plasma and the intrinsic quantum properties of the electron gas itself rather than having an extrinsic cause.

We compare our experimental data to three different models: a purely classical local-response Drude model which assumes a constant electron density profile in the metal nanoparticle, a semiclassical local-response Drude model where the electron density is determined from the quantum mechanical problem of electrons moving in an infinite spherical potential well~\cite{Keller:1993}, and finally, a semiclassical model based on the hydrodynamic description of the motion of the electron gas which takes into account nonlocal response through the internal quantum kinetics of the electron gas in the Thomas--Fermi (TF) approximation~\cite{Bloch:1933a, Boardman:1982a}. We find good qualitative agreement between our experimental data and the two semiclassical models, thus supporting the anticipated nonlocal nature of SPs of Ag nanoparticles in the 1-10~nm size regime. The experimentally observed blueshift is however significantly larger than the predictions by the two semiclassical models.

\section{Materials and Methods}

The nanoparticles are grown chemically following the method described in Ref.~\cite{Mulfinger:2007} and subsequently stabilized in an aqueous solution with borohydride ions. The mean size of the nanoparticles is 12~nm with a very broad size distribution ranging from 3 to 30~nm. The nanoparticle solution is dispersed on a 10~nm thick commercially available $\text{Si}_3\text{N}_4$ membrane (TEMwindows.com), which has a refractive index of approximately $n\approx2.1$~\cite{Baak:1982}. To characterize our nanoparticles we have used an aberration-corrected STEM FEI Titan operated at 120~kV with a probe diameter of approximately 0.5~nm, and convergence and collection angles of 15~mrads and 17~mrads, respectively. The Titan is equipped with a monochromator allowing us to perform EELS with an energy resolution of $0.15\pm0.05$~eV. We systematically performed EELS measurements at the surface and in the middle of each nanoparticle. The EELS spectra were taken with an exposure time of 90~ms to avoid beam damage as much as possible. To improve the signal-to-noise ratio we accumulated ten to fifteen spectra for each measurement point. We observed no evidence of damage after each measurement.

The experimental data were analyzed with the aid of commercially available software (Digital Micrograph) and three different methods were used to reconstruct and remove the zero-loss peak (ZLP): the first method is the reflected tail (RT) method, where the negative-energy half part of the ZLP is reflected about the zero-energy axis to approximate the ZLP at positive energies, while the second method is based on fitting the ZLP to the sum of a Gaussian and a Lorentzian functions. The third method is to pre-record the ZLP prior to each set of EELS measurements. All three methods yielded consistent results.

The energies of the SP resonance peaks were determined by using a nonlinear least-squares fit of our data to Gaussian functions. The error in the resonance energy is given by the 95~\% confidence interval for the estimate of the position of the center of the Gaussian peak. Nanoparticle diameters were determined by calculating the area of the imaged particle and assigning to the area an effective diameter by assuming a perfect circular shape. The error bars in the size therefore correspond to the deviation from the assumption of a circular shape, which is estimated as the difference between the largest and smallest diameter of the particle.

\section{Theory}

In the following theoretical analysis our hypothesis is that the blueshift of the SP resonance energy is related to the properties of the electron density profile in the metal nanoparticle. Therefore, we use three different approaches to model the electron density of the Ag nanoparticle. In all three approaches, we calculate the optical response and thereby also the resonance energies of the nanoparticle through the quasistatic polarizability $\alpha$ of a sphere embedded in a homogeneous background dielectric with permittivity $\epsilon_\text{B}$. With this approach, we make two implicit assumptions: the first is that we can neglect retardation effects and the second is that we can neglect the symmetry-breaking effect of the substrate. We have validated the quasistatic approach by comparing to fully retarded calculations~\cite{Ruppin:1973}, which shows excellent agreement in the particle size range we consider. The effect of the substrate will be taken into account indirectly by determining an effective homogeneous background permittivity $\epsilon_\text{B}$ using the average resonance frequency of the largest particles ($2R>20$~nm) as the classical limit.

The first, and simplest, approach is to assume a constant free-electron density $n_0$ in the metal particle, which drops abruptly to zero outside the particle. This assumption is the starting point of the classical local-response Drude model for the response of the Ag nanoparticle, where the polarizability is given by the Clausius--Mossotti relation, which is well-known to be size independent for subwavelength particles. The classical local-response polarizability $\alpha_\text{L}$ is~\cite{Maier:2007}
\begin{equation}
    \alpha_\text{L}(\omega) = 4 \pi R^3 \frac{\epsilon_\text{D}(\omega)-\epsilon_\text{B}}{\epsilon_\text{D}(\omega)+2\epsilon_\text{B}}, \label{eq:alphaL}
\end{equation}
where $R$ is the radius of the particle and $\epsilon_\text{D}(\omega) = \epsilon_\infty(\omega) - \omega_\text{p}^2/(\omega^2+i\gamma\omega)$ is the classical Drude permittivity taking additional frequency-dependent polarization effects such as interband transitions into account through $\epsilon_\infty(\omega)$, not included in the plasma response of the free-electron gas itself.

The second approach is to correct the standard approximation in local-response theory of a homogeneous electron density profile by using insight from the quantum wave nature of electrons to model the electron density profile and take into account the quantum confinement of the electrons. For nanometer-sized spheres, the classical polarizability given by the Clausius--Mossotti relation must be altered to take into account an inhomogeneous electron density. In Ref.~\cite{Keller:1993}, it is shown that in general the local-response polarizability for a sphere embedded in a homogeneous material is given as
\begin{subequations}
\label{eq:polLQC}
\begin{equation}
    \alpha_\text{LQC}(\omega)=12\pi \int_0^R r^2 \text{d}r \frac{\epsilon(r,\omega)-\epsilon_\text{B}}{\epsilon(r,\omega)+2\epsilon_\text{B}} , \label{eq:alphaLQC}
\end{equation}
now with a spatially varying Drude permittivity~\cite{Ozturk:2011a, Keller:1993}
\begin{equation}
    \epsilon(r,\omega) = \epsilon_\infty(\omega) - \frac{\omega_\text{p}^2}{\omega(\omega+i\gamma)} \frac{n(r)}{n_0}. \label{eq:inhomDrude}
\end{equation}
\end{subequations}
Here, $n(r)$ is the electron density in the metal nanoparticle. Clearly, if $n(r)=n_0$ we arrive at the classical Clausius--Mossotti relation Eq.~(\ref{eq:alphaL}) as expected. To determine the density profile in this local-response model, we follow the approach of Ref.~\cite{Keller:1993} and assume that the free electrons move in an infinite spherical potential well. The approach just outlined of a local-response theory with an inhomogeneous electron density is very similar to the theoretical model used in Ref.~\cite{Scholl:2012} for explaining their experimental results. It should be noted that any effects due to electron spill-out and quantum tunneling are neglected in all of the approaches that we consider.

The third and final approach is to compare our experimental data with a linearized nonlocal hydrodynamic model in which the electron density is allowed to deviate slightly from the constant electron density used in classical local-response theories~\cite{Garcia-de-Abajo:2008, Pendry:2012, Dasgupta:1981, Fuchs:1987}. The dynamics of the electron gas is governed by the semiclassical hydrodynamic equation of motion~\cite{Raza:2011,Toscano:2012,Boardman:1982a}, which results in an inhomogeneous electron density profile. The nonlocal hydrodynamic polarizability $\alpha_\text{NL}(\omega)$ is exactly given as
\begin{subequations}
\label{eq:polNL}
\begin{equation}
    \alpha_\text{NL}(\omega) = 4 \pi R^3 \frac{\epsilon_\text{D}(\omega)-\epsilon_\text{B} \left(1+\delta_\text{NL} \right)}{\epsilon_\text{D}(\omega)+2\epsilon_\text{B} \left(1+\delta_\text{NL} \right)}, \label{eq:alphaNL}
\end{equation}
\begin{equation}
    \delta_\text{NL} = \frac{\epsilon_\text{D}(\omega)-\epsilon_\infty(\omega)}{\epsilon_\infty(\omega)} \frac{j_1(k_\text{L} R)}{k_\text{L} R j_1^\prime(k_\text{L} R)}, \label{eq:deltaNL}
\end{equation}
\end{subequations}
and these results constitute our nonlocal-response generalization of the Clausius--Mossotti relation of classical optics. Here, $k_\text{L} = \sqrt{\omega^2+i\omega\gamma-\omega_\text{p}^2/\epsilon_\infty}/\beta$ is the wave vector of the additional longitudinal wave allowed to be excited in the hydrodynamic nonlocal theory~\cite{Raza:2011,Boardman:1982a}, and $j_1$ is the spherical Bessel function of first order. Finally, within TF theory $\beta^2=3/5\,v_\text{F}^2$, where $v_\text{F}$ is the Fermi velocity~\cite{Boardman:1982a}. We emphasize that for $\beta \rightarrow 0$, the local-response Drude result is retrieved, since $\delta_\text{NL} \rightarrow 0$ and Eq.~(\ref{eq:alphaNL}) simplifies to the classical Clausius--Mossotti relation Eq.~(\ref{eq:alphaL}).

\begin{figure}[t!]
\centering
\includegraphics[width=1\columnwidth]{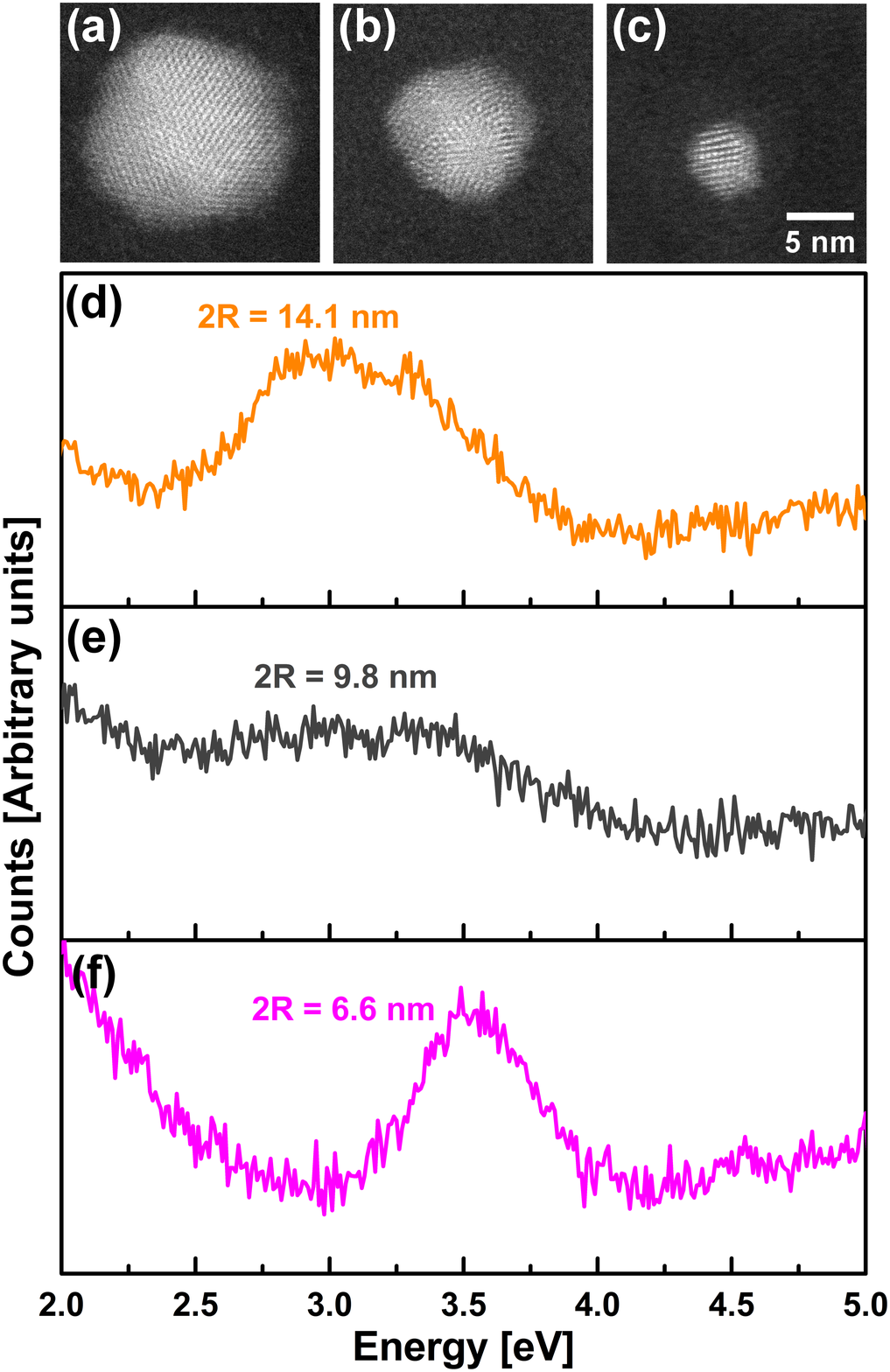}
\caption{Aberration-corrected STEM images of Ag nanoparticles with diameters (a) 15.5~nm, (b) 10~nm, and (c) 5.5~nm, and normalized raw EELS spectra of similar-sized Ag nanoparticles (d-f). The EELS measurements are acquired by directing the electron beam to the surface of the particle.}
\label{fig:fig1}
\end{figure}

The SP resonance energy follows theoretically from the Fr\"{o}hlich condition, \textit{i.e.} we must consider the poles of Eq.~(\ref{eq:alphaNL}). For sufficiently small blueshifts and neglecting damping, the resonance frequency can be approximated by
\begin{equation}
    \omega = \frac{\omega_\text{P}}{\sqrt{\text{Re}[\epsilon_\infty(\omega)]+2\epsilon_\text{B}}} + \sqrt{\frac{2\epsilon_\text{B}}{\text{Re}[\epsilon_\infty(\omega)]}} \frac{\beta}{2R} + O\left(\frac{1}{R^2}\right), \label{eq:resonanceNL}
\end{equation}
where the first term is the common size-independent local-response Drude result for the SP resonance that also follows from Eq.~(\ref{eq:alphaL}), and the second term gives the size-dependent blueshift due to nonlocal corrections. At this stage, we note that a $1/(2R)$ dependence was experimentally observed in Refs.~\cite{Charle:1989,Kreibig:1985} using optical spectroscopy. However, Eq.~(\ref{eq:resonanceNL}) reveals, besides a $1/(2R)$ dependence, that there is a delicate interplay in the blueshift between the material parameters of the metal, through $\epsilon_\infty(\omega)$ and $\beta$, and the background medium $\epsilon_\text{B}$. Furthermore, Eq.~(\ref{eq:resonanceNL}) shows that the blueshift can be enhanced with a large-permittivity background medium.

\section{Results}

Figures~\ref{fig:fig1}(a-c) display STEM images of Ag nanoparticles with diameters of 15.5, 10.0, and 5.5~nm respectively. The images show that no chemical residue is left from the synthesis and that the particles are faceted. We find that approximately 70\% of the studied nanoparticles have a relative size error (\textit{i.e.} the ratio of the size error bar to the particle diameter) below 20\% (determined from the 2D STEM images), verifying that the shape of the nanoparticles is to a first approximation overall spherical (see Supplementary Figure 1). On a subset of the particles, thickness measurements using image recordings at different tilt angles were performed, revealing information about the shape of the nanoparticle in the third dimension. Such 3D investigations confirmed that the shape is overall spherical, but however could not be completed for all particles due to stability issues: the positions of tiny nanoparticles fluctuate under too long exposure of the electron beam, thus preventing accurate determination of the shape of the nanoparticle in the third dimension perpendicular to the substrate.

Figures~\ref{fig:fig1}(d-f) display raw normalized EELS data, acquired on Ag nanoparticles with diameters 14.1, 9.8, and 6.6~nm, respectively.
The peaks correspond to the excitation of the SP. When the diameter of the nanoparticle decreases, the SP resonance clearly shifts progressively to higher energies. Figs.~\ref{fig:fig1}(d-f) also display that the amplitude and linewidth of the SP resonances can vary from particle to particle (with the same size) and at times show narrowing instead of the expected broadening of the resonance for decreasing nanoparticle sizes \cite{Genzel:1975,Kreibig:1985,Kraus:1983}. This is for example seen in the linewidths in Figs.~\ref{fig:fig1}(d-f) which seem to decrease with size. However, as will be explained in more detail in the next paragraph, we did not find a systematic trend of the linewidths in our EELS measurements probably due to the shape variations in our ensemble of nanoparticles.

\begin{figure}[b!]
\centering
\includegraphics[width=1\columnwidth]{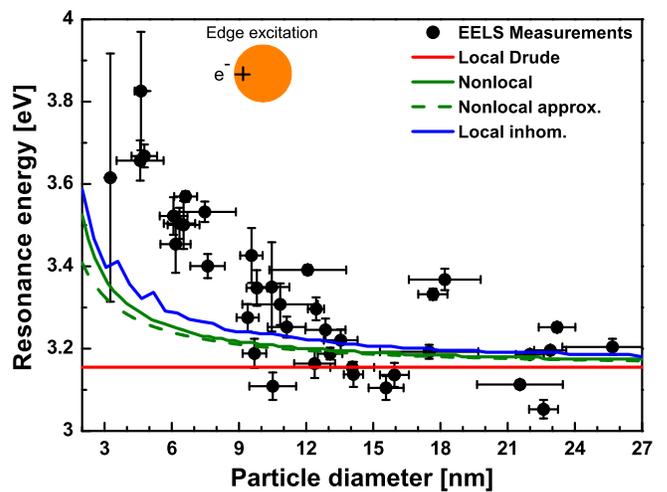}
\caption{Nanoparticle SP resonance energy as a function of the particle diameter. The dots are EELS measurements taken at the surface of the particle and analyzed using the RT method, and the lines are theoretical predictions. We use parameters from Ref. \cite{Rakic:1998a}: $\hbar \omega_\text{p} = 8.282$~eV, $\hbar \gamma=0.048$~eV, $n_0 = 5.9 \times 10^{28}\; \text{m}^{-3}$ and $v_\text{F} = 1.39 \times 10^6$~m/s. From the average large-particle ($2R>20$~nm) resonances we determine $\epsilon_\text{B}=1.53$.}
\label{fig:fig2}
\end{figure}
Figure~\ref{fig:fig2} displays the resonance energy of the SP as a function of the diameter of the nanoparticles. A significant blueshift of the SP resonance of 0.5~eV is observed when the nanoparticle diameter decreases from 26 to 3.5~nm. This trend is in good agreement with the results shown in Ref.~\cite{Scholl:2012}, despite the difference in the substrate and the STEM operating conditions, a strong indication that the blueshift of Ag nanoparticles is robust to extrinsic variations. Another prominent feature in Fig.~\ref{fig:fig2} is the scatter of resonance energies at a fixed particle diameter. We mainly attribute the spread in resonance energies at a given particle size to shape variations in our ensemble of nanoparticles (see Supplementary Material). Slight deviations from perfect circular shape in the STEM images will result in a delicate dependency on the location of the electron probe and give rise to splitting of SP resonance energies due to degeneracy lifting. In this regard, we also note that even a perfectly circular particle on a 2D STEM image may still possess some weak prolate or oblate deformation in the third dimension, resulting in a departure from spherical shape. Calculations using the local response model show that a 20\% deformation of a sphere into an oblate or prolate spheroid results in a $\sim0.4$~eV spread in resonance energy (see Supplementary Figure~2), which is approximately the spread in resonance energy we observe for particles larger than 10~nm. Furthermore, shape deviations may also impact the linewidth of the SP resonance, since the electron probe can excite the closely-spaced non-degenerate resonance energies simultaneously, which may appear as a single broadened peak. This broadening mechanism could explain the apparent linewidth narrowing for decreasing particle size seen in Figs.~\ref{fig:fig1}(d-f). However, we cannot rule out that other effects beyond shape deviations contribute to the spread of resonance energies and impact the SP resonance linewidth. These could for example be the facets or the particle-to-substrate interface~\cite{Noguez:2007}.

Along with the EELS measurements in Fig.~\ref{fig:fig2}, we show Eq.~(\ref{eq:alphaL}) for the local-response Drude model (red line) and the semiclassical local-response model Eq.~(\ref{eq:polLQC}) (blue line).
Furthermore, the nonlocal relation of Eq.~(\ref{eq:polNL}) (green solid line) and the approximate nonlocal relation of Eq.~(\ref{eq:resonanceNL}) (green dashed line) are also depicted, and we see that Eq.~(\ref{eq:resonanceNL}) is accurate for particle sizes $2R \gtrsim 10$~nm.

Due to the narrow energy range in consideration ($\sim~3.0-3.9$~eV), we approximate $\epsilon_\infty(\omega)$ as a second-order Taylor polynomial based on the frequency-dependent values given for Ag in Ref. \cite{Rakic:1998a}. We find $\epsilon_\infty(\omega)=(59.8+i55.1)(\omega/\omega_\text{p})^2 - (40.3+i42.4) (\omega/\omega_\text{p}) + (10.5+i8.6)$. Since the refractive index of the $\text{Si}_3\text{N}_4$ substrate varies hardly ($n\approx 2.1$) in the narrow energy range we consider~\cite{Baak:1982}, we assume that the background permittivity $\epsilon_\text{B}$ is constant and determine it by approximating the average resonance energy of the largest particles ($2R>20$~nm) as the classical limit, \textit{i.e.} the first term of Eq.~(\ref{eq:resonanceNL}).

It is known that local Drude theory produces size-independent resonance frequencies of subwavelength particles, but this theory is clearly inadequate to describe the measurements of Fig.~\ref{fig:fig2}. The nonlocal quasistatic hydrodynamic model predicts a blueshift in agreement with the experimental EELS measurements. Interestingly, the measured blueshift is even larger than predicted. We also see that the local-response model with an inhomogeneous electron density profile shows a similar trend as the nonlocal hydrodynamic model, indicating that these two different models describe very similar physical effects. The oscillations in the resonance energy in the inhomogeneous local-response model seen for small particle diameter are due to small variations in the density profile with decreasing size (\textit{i.e.} discrete changes in the number of electrons), as also stated in Ref.~\cite{Scholl:2012}.

The inhomogeneous local-response model and the nonlocal hydrodynamic model, when applied to a sphere in a homogeneous background medium, agree qualitatively with the EELS measurements. However, they do not provide the full picture. One of the probable issues arising is that the substrate is taken into account indirectly through a homogeneous background medium, a state-of-the-art procedure~\cite{Scholl:2012} which however may not be adequate to describe the effects of the presence of a dielectric substrate. It has been shown that the dielectric substrate modifies the absorption spectrum of an isolated sphere~\cite{Ruppin:1992} and also the waveguiding properties of nanowires~\cite{Nicoletti:2011,Li:2010,Zhang:2012}. In an attempt to include the symmetry breaking effect of the substrate in our theoretical analysis, we apply a simple image charge model. The main effect of the substrate in this picture stems from the interaction of the dipole mode of the nanoparticle with the induced dipole mode in the substrate~\cite{Yamaguchi:1974,Jain:2007, Novotny:2006}. However, we find that such a dipole-dipole model for the substrate is inadequate to describe the large blueshift observed experimentally (see Supplementary Material). Indeed, it has been shown that the induced image charges in the substrate can make the contributions of higher order multipoles in the nanoparticle important~\cite{Ruppin:1983}, and it has also been observed theoretically that higher order multipoles produce larger blueshifts in the nonlocal hydrodynamic model (Fig.~2 in Ref.~\cite{Boardman:1977}). The impact of the substrate on the electron density inhomogeneity and thereby the SP resonance energy depends on the thickness and refractive index of the substrate, which may explain the quantitative agreement between theory and experiment reported in Ref.~\cite{Scholl:2012}, since thinner substrates with smaller refractive indexes were used in their experiments. In order to completely address this issue, one would need to go beyond the dipole-dipole model for the substrate, thus future 3D EELS simulations taking nonlocal effects and/or inhomogeneous electron densities into account would be needed.

Another complementary explanation in the context of the inhomogeneity of the free-electron density could be the combined contribution of both the inhomogeneous static equilibrium electron density and nonlocality. It is well-known that the static equilibrium electron density is inhomogeneous, even in a semi-infinite metal~\cite{Lang:1970}, due to Friedel oscillations and the electron spill-out effect at the metal surface. The Friedel oscillations are modeled in the local quantum-confined model given by Eq.~(\ref{eq:polLQC}) while nonlocality is neglected, and \textit{vice versa} in the nonlocal hydrodynamic model given by Eq.~(\ref{eq:polNL}). As seen in Fig.~\ref{fig:fig2}, the two effects separately give rise to similar-sized blueshifts, suggesting that the contribution of both effects simultaneously could add up to the significantly larger experimentally observed blueshift. Simply put, an extension of the nonlocal hydrodynamic model to include an inhomogeneous equilibrium free-electron density could produce a larger blueshift, which may be in accordance with the experimental observations. Furthermore, such a model could also take into account the electron spill-out effect, which in free-electron models has been argued to produce a redshift of the SP resonance~\cite{Ascarelli:1976a,Boardman:1977,Ruppin:1978a,Apell:1982a,Ljungbert:1985}, describing adequately simple metals. In contrast, it has also been shown that the spill-out effect in combination with the screening from the \textit{d} electrons gives rise to the blueshift seen in Ag nanoparticles~\cite{Liebsch:1993}.

Additional size effects such as changes of the electronic band structure of the smallest nanoparticles, which are considerably more difficult to take into account, also impact the shift in SP resonance energy~\cite{Kreibig:1985}.

\section{Conclusions}
We have investigated the surface plasmon resonance of spherical silver nanoparticles ranging from 26 down to 3.5~nm in size with STEM EELS and observed a significant blueshift of 0.5~eV of the resonance energy. We have compared our experimental data with three different models based on the quasistatic optical polarizability of a sphere embedded in a homogeneous material. Two of the models, a nonlocal hydrodynamic model and a generalized local model, incorporate an inhomogeneity of the electron density induced by the quantum wave nature of the electrons. These two different models produce similar results in the SP resonance energy and describe qualitatively the blueshift observed in our measurements. Although our exact hydrodynamic generalization of the Clausius--Mossotti relation predicts a nonlocal blueshift that grows fast [as $1/(2R)$] when decreasing the diameter and increases even faster for the smallest particles ($2R<10$~nm), the observed blueshifts are nevertheless larger than predicted.

The quantitative agreement between the two different theoretical models and the discrepancy with the larger observed blueshift suggest that a more detailed theoretical description of the system is needed to fully understand the influence of the substrate and the effect of the confinement of free electrons on the SP resonance shift in silver nanoparticles. On the experimental side, further EELS studies of other metallic materials and on different substrates could unveil the mechanism behind the size dependency of the SP resonance of nanometer scale particles.

\emph{Acknowledgments.} We thank S.~I.~Bozhevolnyi for directing our attention to the theoretical model in Ref.~\cite{Keller:1993} and G.~Toscano for fruitful discussions. The Center for Nanostructured Graphene is sponsored by the Danish National Research Foundation, Project DNRF58. The A.~P.~M{\o}ller and Chastine~Mc-Kinney~M{\o}ller Foundation is gratefully acknowledged for the contribution toward the establishment of the Center for Electron Nanoscopy.

\end{document}